\documentstyle[amsfonts,aps,prc,multicol,epsfig]{revtex}

\begin{document}

\draft
\title{Analytical approximation for the sphere-sphere Coulomb potential}

\author{R.Anni}

\address{
Dipartimento di Fisica dell'Universit\`a, Universit\`a di Lecce, Lecce,
Italy\\
Istituto Nazionale di Fisica Nucleare, Sezione di Lecce, Lecce, Italy\\
}

\maketitle
\widetext

\begin{abstract}
%
A simple analytical expression, which closely approximates the Coulomb
potential between two uniformly charged spheres, is presented. This
expression can be used in the optical potential semiclassical analyses
which require that the interaction be analytic on and near the real $r$-axis.
\end{abstract}


\begin{multicols}{2}

\narrowtext

The semiclassical analyses of the optical potential cross sections play an 
important role in understanding the scattering mechanism of light- and 
heavy-ions\cite{BRI85}.
The direct evaluation of the semiclassical amplitudes is based on the
calculation of the integral actions between the turning points of the
radial motion.
The turning points of the radial motion are solution of the equation
\begin{equation}
\label{TP}
k^2(r,\lambda)=k^2 (1-\frac{U(r)}{E}-\frac{\lambda^2}{k^2r^2})=0,
\end{equation}
where $k, \lambda, E,$ and $U(r)$ are, respectively, the wave number,
the angular momentum in units of $\hbar$, the center of mass energy,
and the complex optical potential.
The integral actions are defined by
\begin{equation}
\label{Azi}
S_{ij}(\lambda)=\int_{r_i}^{r_j} k(r,\lambda)\ dr,
\end{equation}
where $r_i$ and $r_j$ are appropriate solutions of Eq.\ (\ref{TP}).

In the semiclassical approximations both these quantities are treated as
analytical functions of $\lambda$ around the real $\lambda$-axis.
This is true only if the complex optical potential is an analytical function
of $r$, on and near the real $r$-axis.

In the phenomenological analyses the optical potential between heavy-ions
is written in the form
\begin{equation}
\label{OptPot}
U(r)=V(r)+i W(r) + V_C(r),
\end{equation}
where $V(r)$ and $W(r)$ are the real and the imaginary part of the
nuclear interaction, and $V_C(r)$ is the Coulomb interaction potential.

As a rule, analytic functions are used for the real and the imaginary part
of the interaction.
On the contrary, the Coulomb part of the interaction is described using
either the Coulomb potential of one point charged particle with a
uniformly charged sphere of radius $R_C$ (in the least recent works) or 
the Coulomb potential between two uniformly charged spheres of radii, say, 
$R_1$ and $R_2$ (in the more recent ones). 
These potentials are not analytic functions around the real $r$-axis.

This fact, rigorously, excludes the possibility of applying the semiclassical 
methods which assumes the analyticity of the potential.
The lack of analyticity must however be considered accidental. 
It arises from our preference to use simple expressions for the 
interaction, and we do not really think that it is connected with some 
physical fact. 
Owing to this we expect that this, and similar troubles, can be cured 
with a little effort. 
 
To do this we here briefly recall some techniques used in the past
to treat the point-sphere case, and we present a similar method for the
sphere-sphere case.
 
Both these Coulomb potentials can be written in the form
\begin{equation}
\label{CouPot}
V_C(r)=\frac{Z_1 Z_2 e^2}{R_C}\ f_C(x),
\end{equation}
where $Z_1$ and $Z_2$ are the charges of the colliding partners, $x=r/R_C$, 
and $f_C(x)$ is an appropriate form factor depending on the case considered. 
For the two sphere case, obviously, $R_C$ is the sum of $R_1$ and $R_2$.
 
The form factor for the point-sphere Coulomb potential is given by
\begin{equation}
\label{pSC}
f_C^{pS}(x)= \left \{ 
\begin{array}{ll}
\frac{1}{2}(3-x^2) & \mbox{$x \leq 1$} \\
\frac{1}{x} & \mbox{$x > 1$.}
\end{array}
\right. 
\end{equation}

This form factor is continuous, as is its first derivative, on the real
$x$-axis, but has a discontinuity at $x=1$ in the second derivative.

To avoid problems connected with this discontinuity, Knoll and 
Schaeffer\cite{KNO76}, in their semiclassical analysis, simply described 
the Coulomb interaction by using the point-point Coulomb potential for all 
the $r$-values. 
This choice is only justified for strongly absorptive interactions, for which the contributions from the internal region of the interaction
are unimportant.
In order to be free from this restriction, Brink and Takigawa \cite{BRI77},
in their uniform semiclassical analysis, preferred to approximate the 
point-sphere form factor given by Eq.\ \ref{pSC} with
\begin{equation}
\label{pSCB}
f_{BT}^{pS}(x)= \frac{1}{x} [1 -
\exp (-a_1 x - a_2 x^2 - a_3 x^3)].
\end{equation}

By imposing the condition that the values of this form factor, and of its 
first two derivatives, coincide with the corresponding quantities of the 
exact one at $x=0$, one obtains $a_1=3/2$, $a_2=9/8$, and $a_3=19/16$. 
In this way the analytical form factor closely approximates the 
point-sphere one, with a maximum relative error smaller than 3\%.
 
A similar analytical form factor, with the additional constraint of being 
an even function of $x$, was also used in the past. 
The evenness of the potential ensures that the scattering functions, of the 
multireflection expansion of the Brink and Takigawa approximation, have 
definite reflection properties in the complex $\lambda$-plane. 
These properties are useful in the semiclassical analysis, and to ensure their 
validity in Ref.\ \cite{ANN81}, although not there explicitly given, the 
following expression was used for the point-sphere form factor
\begin{equation}
\label{pSCAR}
f_{AR}^{pS}(x)= \frac{1}{x} 
\tanh (a_1 x+ a_3 x^3).
\end{equation}

Using the same condition used by Brink and Takigawa one obtains for the
parameters the values $a_1=3/2$, and $a_3=5/8$. With this parameter values, 
the maximum relative error between the approximated and the exact form 
factors is lower than  4\%.

\begin{figure} 
\label{FIG1}
\hspace*{-3mm}
\epsfig{file=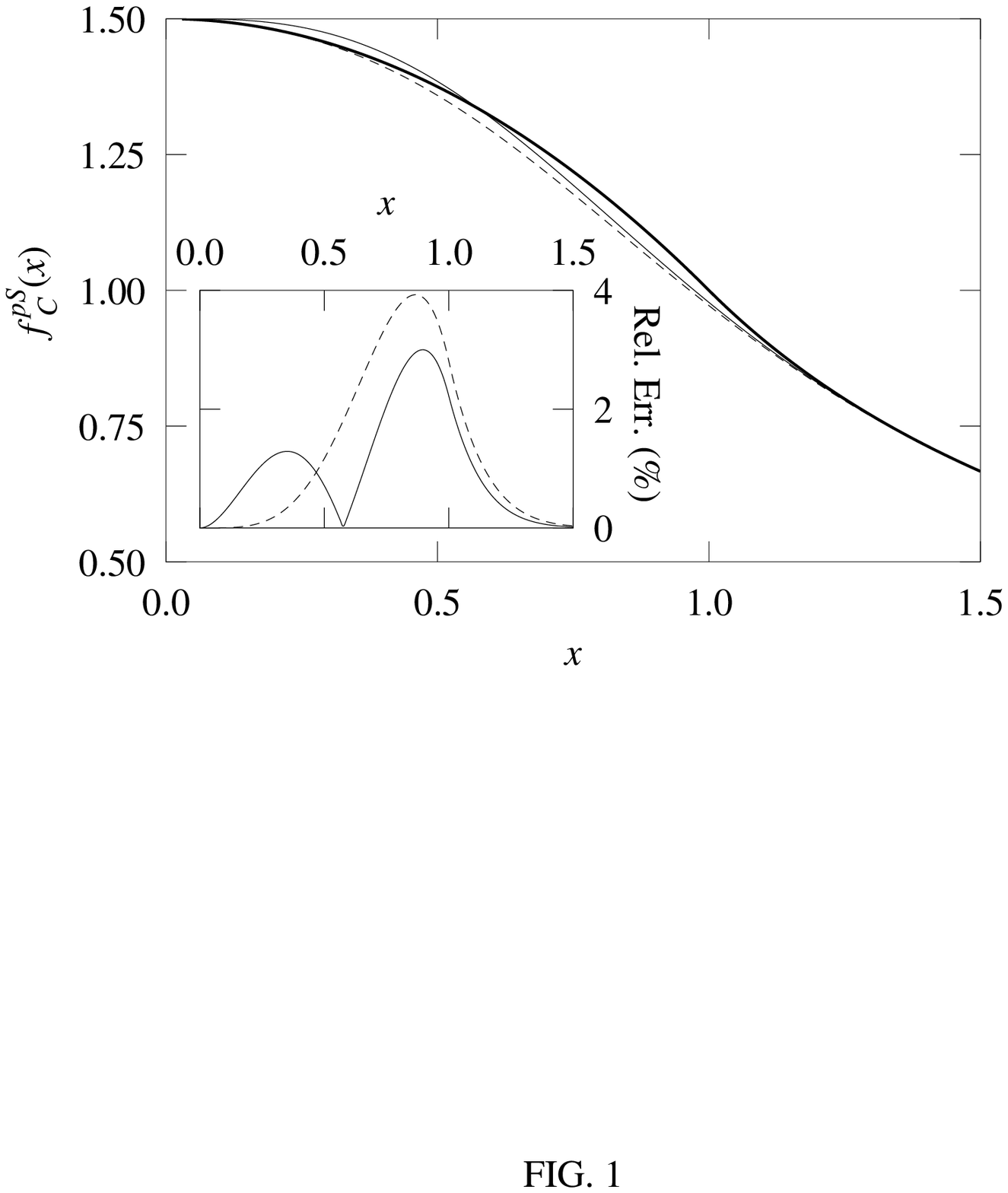,width=8.4cm, clip=}
\caption{The thick curve shows the point-sphere form factor and the thin
solid and dashed lines show the approximations of Eqs.\ (\ref{pSCB}) and 
(\ref{pSCAR}), respectively. The relative errors of the approximations 
are shown in the inset.}  
\end{figure}

In Fig. 1 the behavior of the exact point-sphere form factor is compared
with that of the two above approximations.

The use of the point-sphere Coulomb potential is not, however, 
appropriate to describe realistically the Coulomb interaction in the
heavy-ion scattering processes in which the projectile and the target have
similar radii.
More than 25 years ago\ \cite{DON74,DEV75,IVE75,JAI75,POL76} the role of
more realistic descriptions of the Coulomb interaction between two
heavy-ions was investigated.

The detailed behavior of the Coulomb interaction depends, obviously, on
the properties of the two charge distributions.
In any case the potentials obtained using realistic charge
distributions can be rather well approximated by the Coulomb
potential of two uniformly charged spheres. 
This is the reason why this potential is used even in the more 
recent and  detailed analyses of the elastic scattering 
between light heavy-ions \cite{NIC99,KHO00,NIC00,OGL00}.

The sphere-sphere Coulomb form factor in given by \cite{IVE75}
\begin{equation}
\label{SSC}
f_C^{SS}(x)= \left \{ 
\begin{array}{ll}
\frac{1}{2} b (3-c-b^2 x^2) & \mbox{$x \leq x_0$} \\
\frac{1}{x} \{ 1-3d^2(1-x)^4  &  \, \\
\,\,\,\,\,\times [1-\frac{2}{15}d(1-x)(5+x)]\} & \mbox{$x_0 < x \leq 1$} \\
\frac{1}{x} & \mbox{$x > 1$,}
\end{array}
\right. 
\end{equation}
where $b=1+1/a$, $c=3/5a^2$, $d=(a+1/a+2)/4$, $x_0=(a-1)/(a+1)$,
with $a=R_1/R_2$ and $R_1 \geq R_2$.

This formula is less popular than the equivalent ones given in
Refs. \cite{DON74,JAI75,POL76}, but is here preferred for its simplicity.

The sphere-sphere Coulomb form factor is continuous for real $x$-values
up to the third derivative. The first discontinuities appear in the
fourth derivative at $x=x_0$ and $x=1$.

A very good approximation for the sphere-sphere form factor can
be obtained by using a slight generalization of Eq.\ (\ref {pSCAR})
\begin{equation}
\label{SSCA}
f_{A}^{SS}(r)= \frac{1}{x} 
\tanh (a_1 x+a_3 x^3+a_5 x^5),
\end{equation}
where the parameters $a_1, a_3,$ and $a_5$ depends only on the
ratio $a$ between the larger and the smaller of
the two radii. 
With respect to Eq.\ (\ref{pSCAR}), only the fifth power term
was added in order to preserve the evenness of the form factor.
To give more flexibility to this form factor the parameters are
allowed to be determined, for each value of $a$, by fitting 
Eq.\ (\ref{SSCA}) to points calculated using Eq.\ (\ref{SSC}). 

This technique was used to obtain the sphere-sphere form factor
parameters in a recent semiclassical analysis \cite{ANN01} of
the optical potential scattering of $^{16}$O on $^{12}$C at 
$E_{\text{Lab}}=132$ MeV. 
In the optical potential considered the Coulomb interaction 
was originally described \cite{OGL00} using the potential of two uniform 
charge distributions of radii $R_1=3.54$ fm and $R_2=3.17$ fm. 
With these radii, the values obtained for the parameters are $a_1=2.387, 
a_3=1.071,$ and $a_5=1.683$. 
Using our optical code, the differences between the cross sections 
calculated with the exact and the approximated Coulomb potential are 
completely negligible. 
In fact, calculating the cross sections from $1^\circ$ to $180^\circ$ 
with a step of $0.25^\circ$, the maximum relative error is lower than 0.09\%

The need to use a fitting procedure to obtain the best estimate
of the parameters may be considered a drawback of the approximation (\ref{SSCA}). 
This complication is however unnecessary because the dependence on $\alpha=1/a$ 
of the best fit parameters is rather smooth and, for practical purposes, 
it is sufficient to know the values of the best fit parameters on 
a rather sparse grid of $\alpha$ values.

The reliability of the fit procedure and the possibility of interpolating
the best fit parameter values was tested using 101 $x$ values, equally 
spaced between 0 and 1.5, for 101 values of $\alpha=1/a$, equally spaced 
between 0 and 1.
In the worst case ($\alpha=0$, which corresponds to the point-sphere case) 
the maximum relative error of the best fit form factor is less then 0.8\%.
\begin{figure} 
\hspace*{-3mm}
\epsfig{file=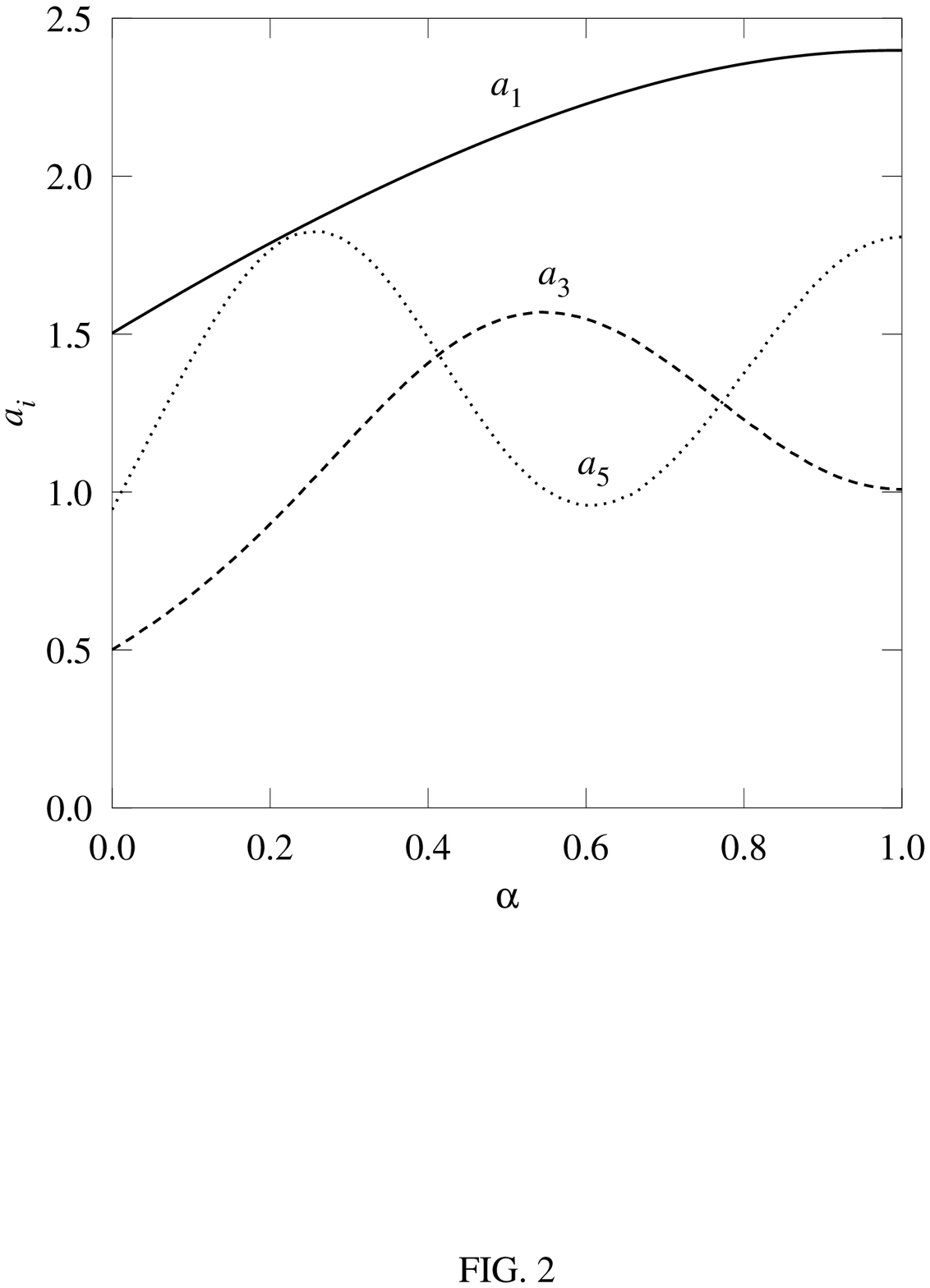,width=8.4cm, clip=}
\caption{Values of the best fit parameters of the sphere-sphere approximated 
form factor as a function of $\alpha$.
}
\label{FIG2}
\end{figure}
\begin{figure} 
\hspace*{-3mm}
\epsfig{file=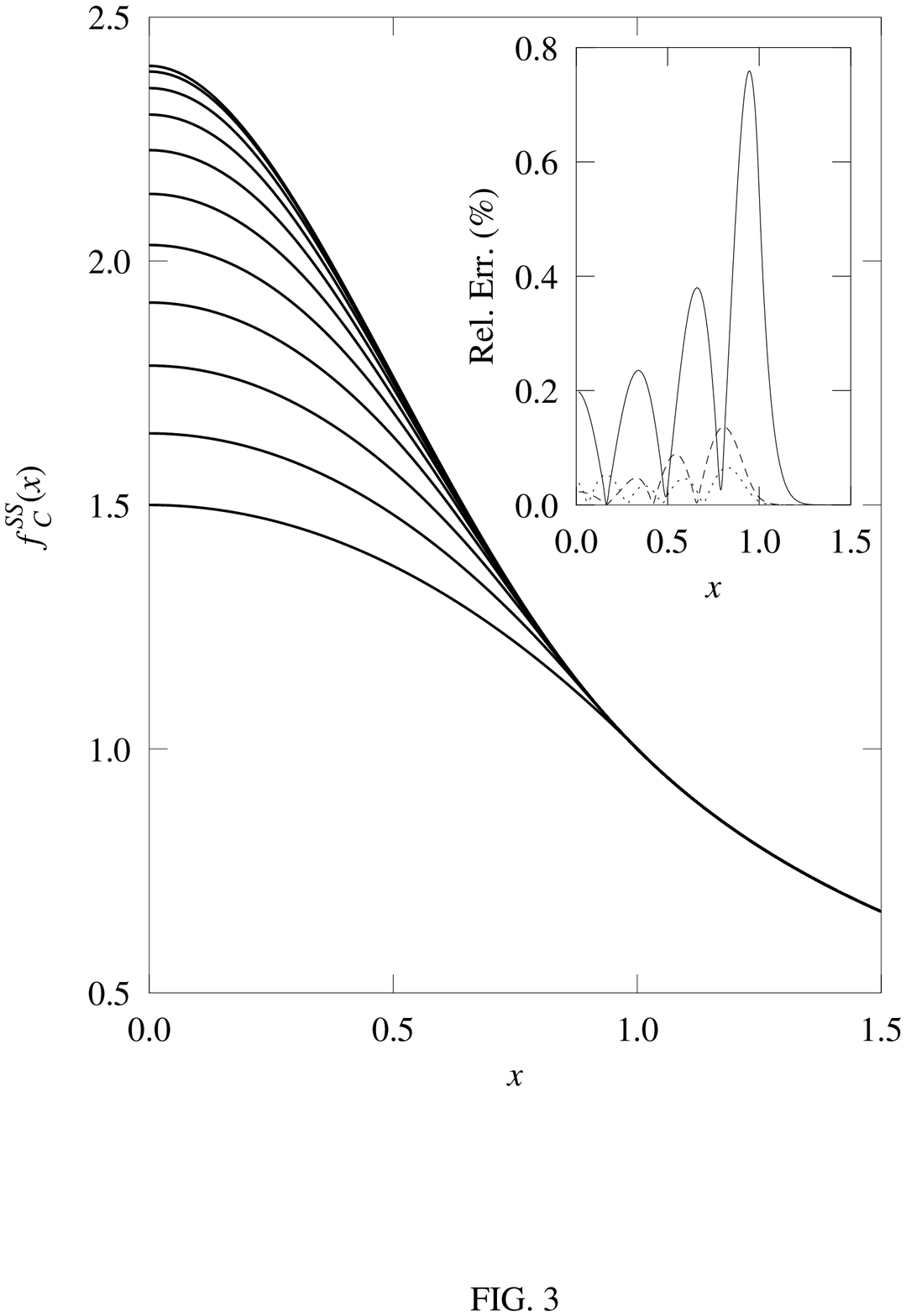,width=8.4cm, clip=}
\caption{Sphere-sphere form factors for $\alpha$ from 0.0 to 1.0, in steps 
of 0.1 (curves from below to above). In the inset the relative errors
for the $\alpha = $ 0.0, 0.5, and 1.0 cases are shown by the solid, dashed, and
dotted lines, respectively. }
\label{FIG3}
\end{figure}
The dependence of the fit parameters on the quantity $\alpha$ is shown 
in Fig.\ \ref{FIG2}.

In Fig.\ \ref{FIG3} we shown the behavior of the exact sphere-sphere form 
factor for the $\alpha$ values from 0.0 to 1.0, with a step of 0.1.
The differences between the exact and the approximated form factors
cannot be appreciated with the thickness used for the curves.
In the inset of the same figure, for the $\alpha=0.0$, $0.5$, and $1.0$ cases,
we show the relative errors between the exact and approximated form factors.

In Tab. \ref{TAB1} we give the values of the best fit parameters
for $\alpha$ values spaced by 0.1. The last column
of the table gives the corresponding maximum relative 
error. Interpolation methods of the tabulated values can be used
to obtain the parameter values at intermediate $\alpha$ values. Using
a linear interpolation at the middle $\alpha$ values the maximum 
relative error is within the bracketing error values for $\alpha<5.0$, 
and does not exceed the 0.21\% for the other $\alpha$ values.

In all the tests that we have performed, the substitution of the sphere-sphere
form factor with its analytical approximation given by Eq.\ (\ref{SSCA}), does not
produce observable effects with the thickness of the lines normally used to plot
an optical cross section. 

In conclusion we can say that the lack of analyticity of the Coulomb 
potentials, traditionally used in the optical model analyses, can be easily
cured without any physically appreciable effect.  
The same is expected to be true also for the, more or less, similar lacks of 
analyticity in other term of the optical interactions, which are sometimes 
used. Using appropriate analytical functions it should be possible to closely
approximate these interactions too. These analytical approximations
could be used in the semiclassical analyses which requires that the
interaction is expressed in terms of analytical functions.

\begin{table}
\begin{tabular}{ccccc}
$\alpha$& $a_1$ & $a_3$ & $a_5$ & Max. Rel. \\
  \,      &  \,     &  \,     & \,      &Err. (\%)\\ 
\tableline
 0.0 &  1.503 &  0.501 &  0.944 &  0.76 \\
 0.1 &  1.650 &  0.674 &  1.420 &  0.60 \\
 0.2 &  1.788 &  0.899 &  1.765 &  0.30 \\
 0.3 &  1.916 &  1.162 &  1.788 &  0.13 \\
 0.4 &  2.033 &  1.408 &  1.487 &  0.11 \\
 0.5 &  2.138 &  1.552 &  1.122 &  0.14 \\
 0.6 &  2.229 &  1.548 &  0.958 &  0.16 \\
 0.7 &  2.302 &  1.416 &  1.077 &  0.15 \\
 0.8 &  2.356 &  1.229 &  1.376 &  0.12 \\
 0.9 &  2.388 &  1.069 &  1.682 &  0.08 \\
 1.0 &  2.399 &  1.008 &  1.808 &  0.07  
\end{tabular}
\caption{Best fit parameters and maximum relative error for
the analytical approximation (\ref{SSCA}) to the 
sphere-sphere Coulomb form factor with different $\alpha$ values.}
\label{TAB1}
\end{table} 

\end{multicols}

\begin{references}

\bibitem{BRI85} D. M. Brink, {\it Semi-Classical Methods for
Nucleus-Nucleus Scattering} (Cambridge University Press), 1985.


\bibitem{KNO76} J.Knoll and R.Schaeffer, 
                Ann. Phys. (N.Y.) {\bf 97}, 307 (1976).

\bibitem{BRI77}  D. Brink and N. Takigawa, 
                 Nucl. Phys. {\bf A279}, 159 (1977).

\bibitem{ANN81} R.Anni and L.Renna,
                il Nuovo Cimento {\bf 65A}, 311 (1981).

\bibitem{DON74} T.W.Donnelly, J.Dubach, and J.D.Walecks,
                Nucl. Phys. {\bf A232}, 355 (1974).

\bibitem{DEV75} R.M. DeVries and M.R.Clover,
                Nucl. Phys. {\bf A243}, 528 (1975).

\bibitem{IVE75} H.Iwe and H.J.Wiebicke,
                JINR-E4-11967 (1978). 

\bibitem{JAI75} A.K.Jain, M.C. Gupta, and C.S.Shastry,
                Phys. Rev. {\bf C12}, 801 (1975).

\bibitem{POL76} J.E.Poling, E.Norbeck, and R.R.Carlson,
                Phys. Rev. {\bf C13}, 648 (1976).


\bibitem{NIC99}  M. P. Nicoli, F. Haas, R. M. Freeman, N. Aissaoui, C. Beck,
A. Elanique, R. Nouicer, S. Szilner, Z. Basrak, A. Morsad, M. E. Brandan,
and G. R. Satchler, Phys. Rev. C {\bf 60}, 064608 (1999).

\bibitem{KHO00}  Dao T. Khoa, W. von Oertzen, H. G. Bohlen, and F. Nuoffer,
Nucl. Phys. {\bf A672}, 387 (2000).

\bibitem{NIC00}  M. P. Nicoli, F. Haas, S. Szilner, Z. Basrak, A. Morsad, G.
R. Satchler, and M. E. Brandan, Phys. Rev. C{\bf 61}, 034609 (2000).

\bibitem{OGL00}  A. A. Ogloblin, Yu. A. Glukhov, W. H. Trzaska, A. S.
Dem'yanova, S. A. Goncharov, R. Julin, S. V. Klebnikov, M. Mutterer, M.
V.Rozhkov, V. P. Rudakov, G. P. Tiorin, Dao T. Khoa, and G. R. Satchler,
Phys. Rev. C {\bf 62}, 044601 (2000).

\bibitem{ANN01} R. Anni, Phys. Rev. C {\bf 63}, 031601 (2001).

\end{references}
\end{document}